\documentclass[12pt]{article}

\newcommand{\xm}{\frac1X}
\newcommand{\corrections}[1]{{#1}}

\title{One dimensional Coulomb-like problem in
deformed space with minimal length}
\author{
T.V. Fityo\footnote{E-mail: fityo@ktf.franko.lviv.ua},\ \ I.O.
Vakarchuk\footnote{E-mail: chair@ktf.franko.lviv.ua}\ \ and V.M.
Tkachuk\footnote{E-mail: tkachuk@ktf.franko.lviv.ua}
\\
  {\small Chair of Theoretical Physics, Ivan Franko National University of Lviv,}\\
{\small 12 Drahomanov St., Lviv, UA-79005, Ukraine
         }}

\begin{document}
\maketitle

\abstract{Spectrum and eigenfunctions in the momentum
representation for 1D Co\-ul\-omb-like potential with deformed
Heisenberg algebra leading to minimal length are found exactly. It
is shown that correction due to the deformation is proportional to
square root of the deformation parameter. We obtain the same
spectrum using Bohr-Sommerfeld quantization condition.}



\section{Introduction}
Quantum mechanics with modification of the usual canonical
commutation relations has attracted a lot of attentions recently.
Such works are motivated by several independent lines of
investigations in string theory and quantum gravity, which suggest
the existence of finite lower bound to the possible resolution of
length $\Delta X$ \cite{Gross88, Maggiore93, Witten96}.

In this paper we consider 1D quantum mechanics with the following
deformation \cite{Kempf94, Kempf95, Kempf97}
\begin{equation}\label{a1}
[X,P]=i\hbar(1+\beta P^2),
\end{equation}
here $\beta P^2$ is a small correction. If $\beta=0$ we obtain
usual algebra. Such a deformation implies that there exists
minimal resolution length $\Delta X\ge\hbar\sqrt\beta$
\cite{Kempf95}, i.e., there is no possibility to measure
coordinate $X$ with accuracy more then $\Delta X$. Note, that
deformed commutation relation (\ref{a1}) gives the same
uncertainty relation as it was suggested in the string theory
\cite{Gross88}. That is why it was assumed that the physical
position and momentum operators could be identified with $X$ and
$P$ operators satisfying deformed commutation relation (\ref{a1}).
Thus, we demand that these operators as well the Hamiltonian $H$
are Hermitian operators.

The use of the deformed commutation relations (\ref{a1}) brings
new difficulties in solving the quantum problems. As far as we
know there are only few problems for which spectra have been found
exactly. They are 1 dimensional oscillator \cite{Kempf95}, D
dimensional isotropic harmonic oscillator \cite{Chang02}, 3
dimensional relativistic Dirac oscillator \cite{Quesne05}. Note,
that in 1 dimensional case the harmonic oscillator problem has
been solved exactly \cite{Quesne03, Quesne04} for more general
deformation leading to nonzero uncertainties in both position and
momentum.

In this paper, we solve the following eigenvalue problem:
\begin{equation}\label{a3}
P^2\psi-\frac\alpha X\psi=E\psi,
\end{equation}
here we put $\hbar=1$ and $2m=1$, $1/X$ means inverse operator of
operator $X$. Recently, in undeformed case this problem has been
considered in \cite{Reyes99, Ran00}. There exists a similar
singular potential $-\alpha/{|X|}$ which had been considered in
detail. Since these potentials are singular there exist a variety
of approaches to the quantization conditions (paper
\cite{Gordeyev97} can be considered as a mini-review on this
topic) and due to different symmetric extensions of operator
$-\alpha/{|X|}$ different spectra are obtained \cite{Tsutsui03}.

Although, the potential $-\alpha/X$ was not studied so intensively
as $-1/|X|$, it has some interesting application in theoretical
physics. In \cite{Moshinsky93} it is shown that this potential
appears in the investigation of mass spectra of mesons
(quark-antiquark systems) in the framework of Dirac oscillators.
The problem, in center-of-mass frame, was reduced to a familiar
radial equation but with the singular potential $V=-b^2/(r^2-a^2)
\sim -b^2/2ax$ where $x=r-a$. The important part of this potential
is just one-dimensional Coulomb-like problem which was studied in
\cite{Moshinsky93}. At the scale of energies and lengths of this
system one can expect appearance of measurable influence of the
minimal length on the energy spectrum. This is one of the main
physical motivation of our studies of the potential $-\alpha/X$ in
deformed space with minimal length.

Also, potential $-\alpha/X$ may have application in the physics of
semiconductors and insulators \cite{Reyes99}. The possibility of
using deformed commutation relation (\ref{a1}) in condensed matter
physics for description of nonpointlike quasi-particles was
pointed out by Kempf \cite{Kempf97b}. In this case minimal length
is interpreted as a free parameter linked with the structure of
nonpointlike particles and their finite size; and no attempt is
made to give an explicit link with some fundamental properties of
the particles \cite{Brau03}. We also would like to present a toy
model which is described by this potential. In this model a point
dipole with constant orientation (constant orientation of the
dipole can be provided by strong uniform electric field) is moving
along a line in the field of an uniformly charged wire which is
perpendicular to the motion line.

In deformed case spectrum of three dimensional hydrogen atom was
approximately found by Brau with the help of perturbation theory
\cite{Brau99}, numerically by Benczik with colleagues
\cite{Benczik05}. Comparison between the ``space curvature''
effects and minimal length effects was made in \cite{Nieto99}. The
authors of \cite{Akhoury03} claimed that they found spectrum, but
it seems that there exists some incorrectness in their paper,
which we discuss below.

The consideration of simple quantum systems in deformed space
gives the possibility to study the influence of deformation on
energy spectra and to find the correction caused by deformation.
Comparing this correction with experimental data it is possible to
estimate the value of deformation parameter (see, for instance,
\cite{Brau99}). Thus, the investigation of quantum mechanical
problems in deformed space is interesting from the mathematical as
well as from the physical points of view.

This paper is organized as follow. In the second section we define
$X$ and $P$ operators and find solutions of the eigenvalue
equation (\ref{a3}). In the third section, we define action of
$1/X$ operator on eigenfunctions and discuss the quantization
condition. In the forth section we obtain spectrum from
Bohr-Sommerfeld quantization condition. And finally, in the fifth
section we discuss our results.

\section{Momentum representation}

There are different representations of algebra (\ref{a1}). One of
them is the so-called quasi-coordinate representation
\cite{Kempf95} $X=x$ and $P=\frac1{\sqrt\beta}\tan\sqrt\beta p$,
where $[x,p]=i$. It is called quasi-coordinate representation
because $X$ does not possess eigenfunctions for which mean value
of kinetic energy is finite and therefore eigenstates of functions
of operator $X$ do not belong to the physical states (for details
see \cite{Kempf95}).

Therefore, we prefer to use the momentum representation. In this
representation momentum and coordinate operators read
\begin{equation}\label{c1}
P=p,\qquad X=i(1+\beta p^2)\frac d{dp}.
\end{equation}
Operator $X$ defined in such a way is an Hermitian operator with
the following definition of scalar product \cite{Kempf95}:
\begin{equation}\label{c2}
\langle\phi|\psi\rangle = \int\limits^\infty_{-\infty} \frac
{\phi^*(p)\psi(p)}{1+\beta p^2} dp.
\end{equation}

For undeformed case the action of inverse operator $1/X$ has been
expressed in the following way \cite{YepezVargas87}
\begin{equation}\label{x1}
\xm\psi(p)=-i\int\limits_{-\infty}^p\psi(q)dq.
\end{equation}
In deformed case we can express it in a similar way
\begin{equation}\label{x2}
\xm\psi(p)=-i\int\limits_{-\infty}^p\frac{\psi(q)}{1+\beta q^2}dq.
\end{equation}
For such a definition $\xm X\psi(p)=X\xm\psi(p)=\psi(p)$. But in
undeformed case the application formula (\ref{x1}) leads to
existence of the only trivial solution $\psi(p)=0$ \cite{Ran00}.
Using the same procedure as in \cite{Ran00} for deformed case we
obtain the same trivial solution.

In order to obtain non-trivial solutions it is necessary to
redefine $1/X$ operator slightly. We rewrite formula (\ref{x2})
\begin{equation}\label{x3}
\xm\psi(p)=-i\int\limits_{-\infty}^p\frac{\psi(q)}{1+\beta q^2}dq
+c,
\end{equation}
where $c$ is a constant. For such a definition $X\xm=1$, but $\xm
X\ne1$. We shall find value of this constant below. Note, that in
undeformed case existence of $c$ in the momentum representation
corresponds to derivative discontinuity of eigenfunction at the
origin in coordinate representation \cite{Ran00}.

Multiplying the eigenvalue equation (\ref{a3}) by $X$ we obtain a
new equation which does not depend on constant $c$.
\begin{equation}\label{f1}
XP^2\psi-\alpha\psi=EX\psi.
\end{equation}
Its explicit form reads
\begin{equation}\label{c4}
i(1+\beta
p^2)\left[p^2\psi'(p)+2p\psi(p)-E\psi'(p)\right]-\alpha\psi(p)=0,
\end{equation}
where $'$ denotes the derivative with respect to $p$. The solution
of the last equation is
\begin{equation}\label{c5}
\psi_\epsilon(p)=\frac{C_\epsilon}{\epsilon+p^2}\exp\left[
\frac{-i\alpha} {1-\epsilon\beta} \left(
\frac1{\sqrt\epsilon}\arctan\frac
p{\sqrt\epsilon}-\sqrt\beta\arctan\sqrt\beta p \right)\right],
\end{equation}
where $\epsilon=-E$ and a normalization constant $C_\epsilon$
reads
$$C_\epsilon=\sqrt{\frac2\pi}\epsilon^{\frac34}\frac
{1+\sqrt{\epsilon\beta}} {\sqrt{1+2\sqrt{\epsilon\beta}}}.$$ If
$\epsilon\le0$ then normalization integral for eigenfunction
(\ref{c5}) $\langle\psi_\epsilon|\psi_\epsilon\rangle$ diverges.
So, we require that $\epsilon>0$.

We can rewrite this eigenfunction in the following way
\begin{equation}\label{c5ss}
\psi_\epsilon(p)=\frac
{C_\epsilon}{\epsilon+p^2}\left(\frac{\sqrt\epsilon+ip}
{\sqrt\epsilon-ip}\right)^{-\frac\alpha{2\sqrt\epsilon(1-\epsilon\beta)}}
\left(\frac{1+i\sqrt\beta p} {1-i\sqrt\beta p}\right)
^{\frac{\alpha\sqrt\beta}{2(1-\epsilon\beta)}}
\end{equation}
In limit $\beta\to 0$ this eigenfunction is equivalent to the
corresponding eigenfunction for undeformed case from paper
\cite{YepezVargas87}.

Function (\ref{c5}) satisfies equation (\ref{f1}) but it does not
satisfy the initial equation (\ref{a3}) with operator $1/X$ in the
form defined by (\ref{x2}). If we define operator $1/X$ in form
(\ref{x3}) we can find such a constant $c$ that eigenfunction
(\ref{c5}) satisfies the eigenvalue equation (\ref{a3}).

The procedure presented in \cite{Akhoury03} leads to correct
expressions of eigenfunctions in one-dimensional case and
eigenfunctions expression from that paper and expression
(\ref{c5ss}) coincide. In \cite{Akhoury03} quantization condition
was chosen from the requirement of single-valuedness of
eigenfunction (\ref{c5ss}). It was stated that this condition is
equivalent to $\frac\alpha{2\sqrt\epsilon(1-\epsilon\beta)}=n$,
but the influence of the term in the second parentheses on the
single-valuedness was neglected. In the next section, we discuss
correct quantization condition.

\section{Spectrum}

To find constant $c$ let us make some manipulations with equation
(\ref{c4}). After the division it by $i(1+\beta p^2)$ and
subsequent integration over $p$ we obtain
\begin{equation}\label{f2}
p^2\psi(p)+i\alpha\int\limits_{-\infty}^p \frac{\psi(q)dq}{1+\beta
q^2}-\alpha c[\psi]=E\psi(p),
\end{equation}
where $c[\psi]$ is an integration constant (with respect to $p$)
being, in general, a functional of $\psi$.

The last equation (\ref{f2}) has the form of the eigenvalue
equation (\ref{a3}). So, we can express action of operator $1/X$
on an eigenfunction as follow
\begin{equation}\label{f4}
\frac1X\psi(p)=-i\int\limits_{-\infty}^p \frac{\psi(q)dq}{1+\beta
q^2}+c[\psi].
\end{equation}
Substituting expression for eigenfunctions (\ref{c5}) into
eigenvalue equation (\ref{f2}) we obtain
\begin{equation}\label{f3}
c[\psi_\epsilon]=\frac1\alpha\lim_{p\to-\infty}(p^2+\epsilon)
\psi_\epsilon(p)=\frac{C_\epsilon}{\alpha}
\exp\left(\frac{i\alpha\pi}
{2(\sqrt\epsilon+\sqrt\beta\epsilon)}\right).
\end{equation}

We require that Hamiltonian corresponding to eigenvalue equation
(\ref{a3}) is an Hermitian operator on its eigenfunction
(\ref{c5}). It is obvious that operator $p^2$ is an Hermitian
operator on these eigenfunctions. Thus, we require that operator
$1/X$ to be an Hermitian operator on the set of eigenfunctions
\begin{equation}\label{f5}
\left\langle\left.\frac1X\psi_{\epsilon_i}\right|\psi_{\epsilon_j}
\right\rangle=\left\langle\psi_{\epsilon_i}\left|\frac1X\right.\psi_{\epsilon_j}
\right\rangle.
\end{equation}
Using expression (\ref{f4}) for operator $1/X$ we can rewrite this
condition
\begin{eqnarray}\label{x5}
i\int\limits_{-\infty}^\infty\frac{\psi_{\epsilon_j}(p)}{1+\beta
p^2}dp \int\limits_{-\infty}^p
\frac{\psi^*_{\epsilon_i}(q)}{1+\beta
q^2}dq+c^*[\psi_{\epsilon_i}]\int\limits_{-\infty}^\infty\frac
{\psi_{\epsilon_j}(p)}{1+\beta p^2}dp=\nonumber\\
-i\int\limits_{-\infty}^\infty\frac{\psi^*_{\epsilon_i}(p)}{1+\beta
p^2}dp \int\limits_{-\infty}^p \frac{\psi_{\epsilon_j}(q)}
{1+\beta q^2}dq+c[\psi_{\epsilon_j}]\int\limits_{-\infty}^
\infty\frac {\psi_{\epsilon_i}^*(p)}{1+\beta p^2}dp.
\end{eqnarray}
According to the facts that
$$
\int\limits_{-\infty}^\infty f(p)dp \int\limits_{-\infty}^p
g(q)dq= \int\limits_{-\infty}^\infty g(p)dp \left[\int\limits_
{-\infty}^\infty f(q)dq -\int\limits_{-\infty}^p f(q)dq\right],
$$
and
\begin{equation}\label{x6}
\int\limits_{-\infty}^\infty \frac{\psi_\epsilon(p)dp}{1+\beta
p^2}=\frac{2C_\epsilon}\alpha \sin g(\epsilon),
\end{equation}
where
\[g(\epsilon)=\frac{\alpha\pi}{2(\sqrt\epsilon+\sqrt\beta\epsilon)},\]
we simplify condition (\ref{x5}) to
\[\sin[g(\epsilon_i)-g(\epsilon_j)]=0.\]
As a result we have $g(\epsilon_i)-g(\epsilon_j)=\pi m$, where $m$
is integer. So, by declaring that some $\epsilon_0$ belongs to the
spectrum we put the following condition on the remaining
eigenvalues $\epsilon$:
\begin{equation}\label{f7p}
\frac{\alpha}{2(\sqrt\epsilon+\sqrt\beta\epsilon)}=
\frac{\alpha}{2(\sqrt\epsilon_0+\sqrt\beta\epsilon_0)}+m
\end{equation}
or
\begin{equation}\label{f7}
\frac{\alpha}{2(\sqrt\epsilon+\sqrt\beta\epsilon)}=\delta+n,
\end{equation}
where $n=m+\left[\frac{\alpha}
{2(\sqrt\epsilon_0+\sqrt\beta\epsilon_0)}\right]$ is integer (here
$[x]$ denotes integer part of $x$), parameter $\delta$ is a
fractional part of $\frac{\alpha} {2(\sqrt\epsilon_0+
\sqrt\beta\epsilon_0)}$ and is in range $0\le\delta<1$.

In fact, we obtain a family of spectra, each of them is
characterized by value of $\delta$. Value of parameter $\delta$
can be calculated from the results of an experiment.

For undeformed case  it is supposed in many papers that
$\delta=0$, it corresponds that eigenfunction vanishes at the
origin ($\psi(x)\big|_{x\to0}=0$). Let us analyze the case of
$\delta=0$ in deformed case more thoroughly. We have
\begin{equation}\label{z1}
\frac{\alpha}{2(\sqrt\epsilon+\sqrt\beta\epsilon)}=n,
\end{equation}
where $n$ is integer. It has no finite solutions for $n=0$ and it
is a quadratic equation in $\sqrt\epsilon$ if $n\ne0$. Its two
real solutions read
\begin{equation}\label{c7}
\sqrt\epsilon_{1,2}=\frac{-1}{2\sqrt\beta} \pm\frac1{2\sqrt\beta}
{\sqrt{1+\frac{2\alpha}n\sqrt\beta} }.
\end{equation}
According to the fact that $\sqrt\epsilon>0$ we have to keep only
one solution with ``$+$'' and, moreover, we have to require that
$n$ and $\alpha$ have the same sign, for $\alpha>0$ we obtain that
$n$ is a positive integer.

So, the spectrum of Hamiltonian (\ref{a3}) for $\delta=0$ is
expressed as follow
\begin{equation}\label{c9}
E_n=-\epsilon=-\frac{1}{4\beta}{\left(1-\sqrt{1+\frac{2\alpha}n\sqrt\beta}\right)^2
},\quad n=1,2,\dots
\end{equation}
For small $\beta$ energy spectrum can be approximated as
\[E_n=
-\frac{\alpha^2}{4n^2}+\frac{\alpha^3}{4n^3} \sqrt\beta -
\frac{5\alpha^4}{16n^4}\beta+o(\beta^{3/2}),\quad n=1,2,\dots\] In
limit $\beta\to0$ the spectrum (\ref{c9}) coincides with the
hydrogen atom spectrum with $\alpha=e^2$, $m=1/2$, $\hbar=1$.

It is interesting to note that if $\delta=0$ then eigenfunction
(\ref{c5}) is a single-valued function of complex variable $p$
except two finite cuts (the first cut goes from $i\sqrt\epsilon$
to $i/\sqrt\beta$, the second one goes from $-i\sqrt\epsilon$ to
$-i/\sqrt\beta$ ). Condition of single-valuedness is widely used
as a quantization criterion \cite{Ran00,Akhoury03,YepezVargas87},
but in our opinion it is an unconvincing criterion.

In general, $\delta\ne0$ and energy spectrum can be obtained in a
similar way and it reads
\begin{equation}\label{c9s}
E_n=-\frac{1}{4\beta}\left(1-\sqrt{1+\frac{2\alpha}{n+\delta}\sqrt\beta}\right)^2
,\quad n=0,1,2,\dots
\end{equation}
Here, due to the fact that $\delta>0$ we obtain one additional
level with quantum number $n=0$. If one tends $\delta$ to zero for
$n=0$ we obtain infinite negative energy and corresponding
eigenfunction vanishes everywhere. Similar properties appear in
undeformed case \cite{Gordeyev97}.

\section{Semiclassical approach}

For 1D undeformed case $[x,p]=i$ and Bohr-Sommerfeld quantization
condition reads
\begin{equation}\label{d1}
2\pi(n+\delta)=\oint pdx,
\end{equation}
where $n$ is an integer, and $\delta$ is a parameter which depends
on the boundary conditions \cite{Landau} ($0\le\delta<1$). In this
section we construct the Bohr-Sommerfeld quantization condition
for deformed case.

Operator $X=i(1+\beta p^2)\frac{d}{dp}$ we rewrite as
\begin{equation}\label{d2}
X=(1+\beta p^2)x,\qquad x=i\frac{d}{dp}.
\end{equation}
Here $x$, $p$ are canonical variables which satisfy $[x,p]=i$.

Classical Hamiltonian corresponding to the system reads
\begin{equation}\label{d3}
H(x,p)=p^2-\frac{\alpha}{(1+\beta p^2)x}.
\end{equation}
From energy conservation law $H(x,p)=E$ we can express $p$ as a
function of $x$ but it is cumbersome to integrate the respective
function. Instead, we use the identity $\oint pdx=-\oint xdp$ and
express $x$ as a function of $p$:
\begin{equation}\label{d4}
x=\frac{\alpha}{(1+\beta p^2)(p^2-E)}.
\end{equation}

When the particle moves from the origin to some turning point
momentum $p$ changes from $+\infty$ to $0$, when the particle
returns to the origin $p$ changes from $0$ to $-\infty$.
Therefore, $-\oint xdp=\int\limits^\infty_{-\infty}xdp$ and
\begin{equation}\label{d5}
2\pi(n+\delta)=\int\limits^{\infty}_{-\infty}\frac{\alpha}
{(1+\beta p^2) (p^2-E)}
dp=\frac{\pi\alpha}{\sqrt{\epsilon}+\epsilon\sqrt\beta},
\end{equation}
here we take into account that for bound states energy
$E=-\epsilon$ is negative.

Condition (\ref{d5}) coincides with condition (\ref{f7}) and we
recover the same spectrum as in formula (\ref{c9s}).

\section{Discussions}

In this section we discuss two interesting properties of the
problem considered in this paper and Bohr-Sommerfeld quantization
condition.

Existence of different spectral families means that there exist
different extensions of operator $1/X$. We think that from the
physical point of view it signifies that we can approximate a
singular potential with different regular ones and as a result we
obtain different spectra (cf. \cite{Gordeyev97}). As one can see,
the existence of minimal length does not remove the singularity of
potential $1/X$.

Another interesting property is that the first correction to the
energy spectrum is proportional to $\sqrt\beta$. It brings an
opportunity to reveal the existence of deformed commutation
relations (\ref{a1}) for smaller $\beta$. For previously solved
problems (harmonic oscillator \cite{Kempf97b}, hydrogen atom
\cite{Brau99}) the correction is proportional to $\beta$.
\corrections{On the other hand, it is difficult to say if it is
possible, using present experimental setup, to find corrections to
energy spectrum caused by deformation of space with minimal length
in quantum systems described by 1D potential $1/X$.}

We also derive energy spectrum using Bohr-Sommerfeld quantization
condition. It is interesting to note that semiclassical result
coincides with the exact result obtained from Schr\"odinger
equation.

\section{Acknowledgment}

We thank Dr.\ Andrij Rovenchak and Yuri Krynytskyi for helpful
discussions. We also would like to thank to Dr.\ Andrij Rovenchak
for careful reading  the manuscript.

\end{document}